\documentclass[a4paper,11pt]{article}
\usepackage{fullpage}
\usepackage{setspace}
\usepackage[T1]{fontenc}
\usepackage[utf8]{inputenc}
\usepackage{lmodern}
\usepackage[british,english]{babel}
\usepackage{datetime}
\usepackage[parfill]{parskip}
\usepackage{underscore}
\usepackage{enumitem}
\usepackage{listings}
\begin{document}

\title{Integrating Communications and Merging Messaging via the eXtensible Messaging and Presence Protocol}
\author{Martin A. COLEMAN\\
Sydney, Australia\\
Email: martin.coleman@monash.edu
}
\date{}

\maketitle
\thispagestyle{empty}
\begin{abstract}
Common problems affecting modern email usage include spam, lack of sender verification, lack of built-in security and lack of message integrity. This paper looks at how we can utilise the extensible messaging and presence protocol also known as XMPP to, in time, replace email facilities. We present several methods for initiating a transition away from SMTP for email to rely upon the inherent benefits of XMPP with minimal disruption to existing networks and email infrastructure. We look at how a program might be used to open an existing POP3/IMAP account, scan for messages that can be sent to a XMPP network user, extract the message and then deliver it the XMPP user's client. We show that the system can be implemented and then deployed with a minimum of hassle and network disruption to demonstrate XMPP as a reliable and fast replacement for email as we know it today.

\end{abstract}

\noindent{\bf Index Terms}: XMPP, email, instant messaging, integration, networks, secure communications, transition, jabber.

\null
\vfill
\thispagestyle{empty}
\begin{flushbottom}
\begin{center}
Copyright \copyright\ 2013 Martin A. COLEMAN. This document may be freely shared, copied, transferred and/or re-distributed, in part or in whole, for any purpose and by any means, provided that credit is given.

Date: 2018.02.25 \\
Permanent ID of this document: e90211c086cd29f8f06cf98b0b48d942.
\end{center}
\end{flushbottom}

\pagebreak

\twocolumn

\section{Introduction}

As everyday communications becomes more and more critical, people are relying on their idea and their words is expressed clearly, quickly and with integrity. But the introduction of any noise may threaten the integrity of the message, in fact it may even destroy the message altogether. So certain safe guards for the message should be introduced.

Email is the world's most popular way to communicate, be in personal messages or business correspondence. However, it is prone to spam, there is no sender verification, there is no receiver verification and you have to set up systems in order to receive them, not to mention they can be delayed by network problems.

Pretty Good Privacy (P.G.P.) was an attempt to introduce sender and receiver verification and authentication, but required a bit of technical knowledge that was out of the reach of the average user and it needed to be configured on both sides in order for it to work. Domain Name Service (D.N.S.) backups and secondary mail servers and mail queues were set up to compensate for network outages and distribution problems. Internet Message Access Protocol (I.M.A.P.) was introduced to complement the backup mail systems so that a message could be reached anywhere and amongst different clients. This was more effective than the other popular system called Post Office Protocol 3 (P.O.P. 3 or POP3).

Unfortunately, even under great operating conditions, the above set up still had flaws and was still prone to spam. Smart software with auto adjusting mail filters were implemented, but message access was still a problem, and prone to excess data usage using clients that fetched or "pulled" a message, rather than having it pushed to them, like an SMS (Short Message Service).

The eXtensible Message and Prescence Protocol (X.M.P.P.) introduces a new messaging foundation that solves all the above problems and bring many other great benefits with it as well that are advantageous to communications.

In this paper, we explore just what XMPP can do for us in regards to reliable and secure communications, how to introduce it in a friendly way to users and existing network infrastructure, deploy it smoothly and then look at potential and viable paths aiming to a full transition from the legacy email systems we use today to a fully XMPP-powered communications backbone for tomorrow.

\section{Evaluation}

\subsection{What Is XMPP?}
The eXtensible Messaging and Presence Protocol (XMPP) is a communications protocol for message-oriented systems and is based on eXtensible Markup Language (XML) \cite{cite01}. Originally known as Jabber it is developed by an open source community to provide a platform for near real-time, instant messaging (IM), contact list maintenance and presence information. With flexibility and extensibility in mind, it has been able to provide a message and notification platform for social networking services, smart grid devices, gaming, file transfer, signalling VoIP as well as many other important and everyday common and practical applications.

Intentional strengths of the system include, but are not limited to:
\begin{itemize}[noitemsep]
\item Decentralisation - Anyone can run their own server. You can even inter-connect with servers run by other people to create a huge network of collaborative XMPP servers.
\item Open Standards - Anyone can create a client or server software suite to use it. No lock-in to what any particular person or company wants you to do. By following the published standard(s) \cite{cite02} anyone can create any type of XMPP integration they like, royalty free.
\item History - XMPP has been around since 1999 \cite{cite03}. Multiple implementations of clients, servers, add-ons, etc exist and have existed since it started.
\item Security - Servers can be made public if you want to run an open network for people to join. You can also run private ones for inside companies for quick, effective and secure networks. You can also run a hybrid of both open and public and private and closed within the same configuration. Strong, secure authentication and message encryption are one of the core set of features, so you can feel good that noone is listening in on your conversation.
\item Flexible - Group chat, network management, file sharing, gaming, remote system monitoring and control, cloud computing, signalling VoIP, identity services, content syndication, the list goes on as to what is possible with XMPP.
\end{itemize}

It's weaknesses are so few and restricted that it takes one line to describe them: data sharing. It is currently inefficient to send large data transfers through due to needing both parties agreeing to the transaction at the same time and no offline saving of data attachments.
However, the majority of people would still use email to send attachments, web-based file sharing facilities such as Dropbox, ownCloud, netMist, etc or even plain FTP for bigger files to send when you are not sure if the other person is online or not.

Beyond the above, it also provides other great and practical benefits. If XMPP were to replace email, as this paper proposes, what does it offer that area beyond any obvious details described above? Exactly how can it match and then extend upon today's common email standard?

Anti-Spam. XMPP will help fight against spam aka unsolicited commercial email. Noone can send you a message without your permission and previous agreement to send/receive messages. It just goes nowhere. Better than that actually, it is not even sent to begin with.

Built-in address book. Every person you have previously agreed to receive messages from is already in your address book, so you can tell at a glance who has your details and is capable of sending you a message. Someone can send through an authorisation request to be added to your address book (or "Roster" as XMPP calls it \cite{cite04} ), but you can ignore it if you like, or see who sent it and dismiss it, completely without that persons knowledge.

Push email. People loved the Blackberry because of this, and XMPP has it built-in. You do not request messages, they are already there. With current email systems such as the Post Office Protocol (POP3), you need to specify the period of time after which the client will re-check to see if there are any new messages. IMAP gets you one step closer to this by sustaining an open connection with the server but it can use up data where data can be scarce, such as on 3G and GSM mobile data systems. XMPP will get that message to your client and/or device without you having to do anything at all as feature for convenience. The sender presses send and you have it in your inbox and/or client near instantly.

Built-in security\cite{cite05}. SMTP is an unauthenticated protocol: with the base protocol you cannot guarantee the identity of who sent a message, nor who the receiver is or will be. Any email can pretend and/or give the impression that it was sent by anyone at all, with no verification. Pretty Good Privacy (PGP) is a great step to offering user and receiver verification, but the protocol was not designed with it in mind, therefore it is not an included feature by default in any client or system.

A POP3/IMAP to XMPP translator/converter makes transitioning to a XMPP-based network very easy and effective. Using current common place protocols that everyone can gain access to and having it automatically be copied onto the newer, faster and more secure network is not only extremely effective, practical and very useful, it is what we delve into with this paper.

Offline messages\cite{cite06}. This is the feature parity that really helps sell the benefits of XMPP. The above benefits all emphasise what can be done with "always connected" clients. But if you suddenly lose the connection to the XMPP server and someone sends you a message, the server holds on to it for you until it sees you re-connect and come back online.

Various clients. Just like email, whether you want to run a desktop client or access one through your web browser like webmail services offer, XMPP gives you that ability. From desktop, to web, to tablets, to smart phones, to your refrigerator and your toaster, XMPP is accessible from them all.

Now that we know what XMPP is and why it is so important to this paper and integrating communications in general, we will look at some of the options available for protocol integration into existing and new networks and we can get to wide deployment of XMPP immediately.

\section{Investigating Some Options}
\indent To merge communications to a XMPP-based foundation, the protocol and associated systems must first gain acceptance, thereby leading to common usage. This might be done by setting up network applications such as instant messengers and demonstrate the benefits of the protocol then build upon that. That way they can see it working and become comfortable with it.

This would inevitably lead to the question of email and using XMPP alongside email, or getting email messages via XMPP until all common correspondents are also on the XMPP network. Having everyone using XMPP would be an admirable goal, indeed it is the one in which this paper amongst a few others, hopes to inspire network administrators to aim for, but for now, integrating current tools into or alongside the new proposed system will hopefully maximise the chances of user acceptance.

\noindent Various ways that this can be accomplished are:
\begin{itemize}[noitemsep]
\item An POP3/IMAP to XMPP account checker.
\item POP3/IMAP to XMPP processor
\item SMTP to XMPP integrated server
\item Plus a few others which derive from the above 3 solutions.
\end{itemize}

Let us examine the above 3 options in more detail.

\subsection{POP3/IMAP to XMPP Account Checker}
A POP3/IMAP to XMPP account checker would run as a program linked with an IMAP library such as libetpan and an XMPP library such as libstrophe. Run by itself or as a cron job, it would either:
Type 1: Connect to a dedicated POP3 or IMAP account, check for messages that are new, parse the headers, retrieve the intended XMPP recipient from a specially formatted subject line, connect via a special XMPP account on a Shared Roster and send the contents of that message through, then continue on through the mail folder, repeating those steps as necessary.
Type 2: Connect to a specified POP3 account that belongs to the XMPP user, check for new messages, parse the headers, retrieve the message subject and content, connect via a special alerts and messages XMPP account either on a Shared Roster or another specially created one just for this purpose and then send that message through, repeating the previous steps as necessary.
The above two operations could even be eventually implemented onto the one program and have it use the necessary method when needed, either as a compile-time switch or as part of a configuration file.

\noindent The Advantages of the above solution are:
\begin{itemize}[noitemsep]
\item Very easy and quick to set up.
\item Easy to integrate.
\item No-one has to even know it is running unless they are on the XMPP network.
\item Requires no changes to be made to any core or central infrastructure.
\item Works perfectly alongside the existing email infrastructure.
\item Takes advantage of offline messages.
\end{itemize}

\noindent The Disadvantages of it are:
\begin{itemize}[noitemsep]
\item Type 2: Prone to spam.
\item Type 1: Practically impossible to spam.
\item Only benefits those who are on the XMPP network.
\item No easy way to respond directly to message sender.
\item Still requires a POP3/IMAP account to work.
\item Does not support attachments.
\end{itemize}

\subsection{SMTP to XMPP Processor}
The SMTP to XMPP processor would be a program that knows how to parse email data and is linked to an XMPP library. It would run on demand in a similar way that procmail is run by postfix as part of the mailbox_command setting. It would take information given to it, then send the information to the intended XMPP recipient.

\noindent The advantages of this solution are:
\begin{itemize}[noitemsep]
\item Easy to set up.
\item No POP3/IMAP account needed.
\item It is handled instantly.
\item Takes advantage of offline messages.
\end{itemize}

\noindent The disadvantages of this method are:
\begin{itemize}[noitemsep]
\item Prone to spam.
\item Does not easily support attachments.
\end{itemize}

\subsection{XMPP Server with Integrated SMTP Server}
XMPP server with integrated SMTP server. This is the Holy Grail of XMPP and SMTP integration. It is the one which will also take the most amount of work to get functional. It is a complete SMTP server and XMPP server and can send, receive, exchange, etc messages between the two protocols under the basic guide of "message passing".

\noindent The advantages of this would be:
\begin{itemize}[noitemsep]
\item Easy to set up.
\item Easy to integrate.
\item Complete, coherent and combined messaging facility.
\item Fast.
\item Switch between the two protocols effortlessly.
\item Supports direct responding to message senders.
\item Roster becomes the Address Book, which becomes the Whitelist.
\item Supports attachments.
\end{itemize}

\noindent The disadvantages of it would be:
\begin{itemize}[noitemsep]
\item Wouldn't encourage users to move away completely from unauthenticated and insecure SMTP as they would see good legacy integration.
\item May be overwhelming to manage Roster or "Address Book" requests without being able to avoid spam attempts.
\end{itemize}

\subsection{Investigation Complete}
So while many individuals and organisations are still quite understandably attached to SMTP and email at the moment, XMPP is a brilliant protocol to introduce into any sized network and there are ways around having both work so that users may benefit from the legacy support of SMTP while taking advantage of the benefits that XMPP brings to modern messaging.

We are still aiming to produce a system that is secure, offers message encryption natively, uses push methods for message receiving, is really, really fast and allows offline messages to be saved. XMPP with the various add-on protocols can definitely get there.

At this point in time, though the XMPP-SMTP integrated server is a good goal to aim for, the simplest, quickest and easiest way to introduce XMPP to a network while still utilising SMTP would be to create the POP3/IMAP to XMPP account checker. Nothing gets in the way to implement it (you just set up a server and tell the interested parties to connect to it), can be easily demonstrated without affecting any other part of the network (no need to disrupt anyone anywhere) and would be suitable for any number of users (limited only by the bandwidth available, which with gigabit networking and compression enabled, is a substantial amount).

This would run when called (either manually or via a cron job on Linux/*BSD) check an email account, process the list of messages for subjects that it recognises and passes the messages on via a XMPP account that is on a Shared Roster and will be received via any XMPP client that the user is currently logged in to.

The next section will detail how I went about making this program and explain its operations.

\section{Implementation}

Of the possible solutions explored in the previous chapter, a software program that connects to a POP3 or IMAP account, checks the subject lines for a certain string of text, retrieves the entire message content of any that matches the criteria, connects to an XMPP server and sends the specified email content to the XMPP user, referred to as the POP3/IMAP to XMPP Account Checker was deemed to be a practical compromise between making no core changes to a network, easy and quick to set up/deploy and great to test with.

First of all, a software library needed to be found that supported POP3 and/or IMAP parsing and access. On top of that, it also needed to have a license that permitted open use in any environment so as to gain the widest acceptance. The libetpan library was chosen as it was available under a BSD 3-clause (no endorsement) license which is friendly towards commercial and non-commercial use.

Second, a library was needed that could talk to the XMPP server, was actively kept up to date with the latest open standards developments happening in the XMPP field and again, was friendly towards linking with another program that could be used in closed, as well as open, computing environments. The libstrophe library was chosen as it was dual licensed under the GPL version 3 (a little restrictive for establishing open standards and reference implementations) and the MIT (very friendly to all environments with practically no restrictions on use).

Third, we need the service itself that will provide and manage the XMPP protocol. In this demonstration ejabberd was chosen as it is a readily available, free, open source and scaleable XMPP server which can run on a variety of operating system platforms. As ejabberd is also a very scaleable server, you will always be able to expand operations to other servers collaboratively once the demonstration proves itself to the network users.

Seeing as I wanted to release my program under the BSD 2-clause license, I needed libraries which were going to be friendly towards close integration and linking to be used in any computing environment; commercial, educational or personal. They also should not be requiring the source code of derivative works to be made available. Any source code changes, fixes, improvements, new features are always greatly appreciated of course, but they are not required if it borders upon becoming a hassle to do.

Now, onto implementing the POP3/IMAP to XMPP Account Checker solution.

\subsection{libstrophe}

libstrophe is a lightweight XMPP client library \cite{cite07} which has minimal dependencies and is configurable for various types of environments. It can be built and used on Unix, Linux and Windows platforms.

libstrophe comes in two forms; first is a library for C for using in circumstances such as standalone software, such as I was, and secondly as a Javascript library. The Javascript version needs a local data and URL fetcher for http_bind, so it needed extra effort to make work and may be a useful foundation for a web-based client, but we are looking at native clients here, so the C version is more than adequate.

From the basics, the library is initialised using xmpp_initialize(), a connection context created using xmpp_ctx_new(), the connection is prepared using xmpp_conn_new(), the connection username is set with xmpp_conn_set_jid(), likewise the password with xmpp_conn_set_pass(), the client creates the TCP connection itself via xmpp_connect_client(), is sustained and monitored via xmpp_run(), then when triggered it is freed up with xmpp_conn_release(), the connection is dropped with xmpp_ctx_free() and then we say goodbye to libstrophe with xmpp_shutdown(). Those are all the essential steps to creating, making and then closing a connection. But we want to send some data now.

A connection handler is set up with xmpp_connect_client which handles software signals such as XMPP_CONN_CONNECT, XMPP_CONN_DISCONNECT and XMPP_CONN_FAIL. This is where we will connect to the email server, browse the inbox looking for special subject lines and send the contents of those messages.

A simple message sending routine might be:

\begin{lstlisting}
void send_message(xmpp_conn_t * const conn, char *message, char *to_jid, void * const userdata)
{
    xmpp_stanza_t *reply, *body, *text;
    xmpp_ctx_t *ctx = (xmpp_ctx_t*)userdata;

    reply = xmpp_stanza_new(ctx);
    xmpp_stanza_set_name(reply, "message");
    xmpp_stanza_set_type(reply, "normal");
    xmpp_stanza_set_attribute(reply, "to", to_jid);
    xmpp_stanza_set_attribute(reply, "xmlns", "jabber:client");

    body = xmpp_stanza_new(ctx);
    xmpp_stanza_set_name(body, "body");

   	text = xmpp_stanza_new(ctx);
    xmpp_stanza_set_text(text, message);
    xmpp_stanza_add_child(body, text);
    xmpp_stanza_add_child(reply, body);

    xmpp_send(conn, reply);
    xmpp_stanza_release(reply);

    return;
}
\end{lstlisting}
this routine is the key function in sending those messages to our users. to_jid is the Jabber/XMPP user ID (XMPP usernames take the same form as email addresses eg: user@server), message is the content of the email itself. This can be any length, but realistically we are limited by the amount of memory allowed to it by the operating system, which is usually enough to send an entire text-based encyclopedia through many times over at once.

We allow some basic error checking but also take advantage of the fact that we are using TCP/IP which is a connection oriented protocol, so making and sustaining a connection is it's job.

The rest is then handled by the POP3/IMAP library.

One impression that one might gain from using the libstrophe library is that it is very signal and messaging (no pun intended) oriented. Just like Win32 programming deals with signals such as WM_RESIZE and WM_PAINT which you write code to deal with, libstrophe uses signals and signal handlers to deal with messages and messaging. C code using libstrophe can be every bit procedural until you start handling the basics of XMPP operations. This makes sense though considering the basic options are like threading; one thread of a program might be a simple Graphical User Interface (GUI) component, while in the background it is keeping time for alerts or even acting on network commands.

So once you set up handlers for the signals you might get, getting code working that deals with what is going on is relatively simple.

\subsection{libetpan}

libetpan is a library containing a set of libraries which provide (e)mail manipulation services contained within a portable and efficient framework \cite{cite08}. It supports IMAP, POP3, SMTP and NNTP.

It natively supports a featured API for the C language and provides numerous examples for interacting with an IMAP/POP3 server.

From the start, a connection is made via init_storage(), to prepare the place to which message data will be gathered. Then we connect using mailfolder_connect(), this is typically the account's INBOX folder. After that, we want to process the INBOX folder and gather information aboutt the messages contained there, especially looking out for a particularly formatted Subject line, so we can determine if it is a normal email message or whether it is intended for out XMPP user.

Next we get a list of messages available via mailsession_get_messages_list(). This gives us an overview of the INBOX folder and can get us started on scanning the subject lines. mailsession_get_envelopes_list() gets us the content of the various fields contained within the emails themselves.

We then scan the subject lines for the string "USER:" which would be followed by the XMPP username or "Jabber ID", grab the content of that message and send it along to the XMPP user.

After all that is done, we clean up after ourselves with mailmessage_list_free(), mailfolder_free() and mailstorage_free() as we want to release that memory and the associated resources for next time.

We then go back to the main XMPP connection thread which will proceed to disconnect us and stop program execution.

\subsection{ejabberd}

ejabberd is a popular XMPP application server which is written mostly in the Erlang programming language \cite{cite09}. Licensed under the GNU General Public License, it is available under numerous operating platforms such as Linux, FreeBSD, NetBSD, OpenBSD, Mac OS X, OpenSolaris and even Microsoft Windows. ejabberd, which stands for Erlang Jabber Daemon (Jabber is what XMPP was called before being renamed due to Jabber being trademarked) is a highly scaleable server which supports clustering and is one of the most popular open source programs written in Erlang.

Using the Debian Linux distribution as an example, the ejabberd configuration file can be found at /etc/ejabberd/ejabberd.cfg. We will modify this file for the very last step. Under "modules", we enable the shared roster feature by uncommenting the mod_shared_roster line. Once that is done, we restart ejabberd via /etc/init.d/ejabberd restart either via sudo or as the root user, then create an account that will provide the alerts service with "ejabberdctl register alerts [domain] [password]".

Then we log in the ejabberd web interface at http://[host]:5280/admin (an admin user presumably is already configured), log in as an admin user, create a shared roster containing the new alerts@[domain] user and the process is complete.

\subsection{Implementation Complete}

Now that we understand how the Account Checker program is put together, and how we can enable an alerts service or user on the XMPP network, we place those XMPP login details and the IMAP user credentials into the source code of the server, compile it and decide how to run it.

This provides the foundation for an SMTP (Email) to XMPP (messaging) transition without any more infrastructure changes than simply running a program and creating a new user on a hopefully already existing or freshly created XMPP network service.

The remaining obstacles beyond network integration, user and widespread usage and acceptance may be specially formatted messages, i.e.: RTF and HTML messages, and email attachments.

Email attachments may be solved via Uniform Resource Indicator (URI) linking scheme, where HTML messages can be turned into a compromise between in-browser rendering of the HTML or enabling a HTML renderer inside a new messaging/email client.

This may be assisted by the development of a complete web-based account portal featuring an optimised XMPP feature, or a web-based service equivalent to webmail but using the XMPP protocol instead, complete with messaging, calendar and address book (roster).

\section{Conclusion}

Communication is an ever changing, ever updating and ever evolving means of getting an idea or expression from one place to another in the easiest, quickest and most efficient way possible.

With every new transition of technology, every new phase of thought with mankind, we have always found ways to use it to communicate better. First it was words, so human language facilitated this. Second, people made marks wherever they could be made permanent, so drawings came about \cite{cite10}. Third came the written word, so more detail could be included without losing accuracy of the detail and the signal had little noise, beyond simple time wearing out the writing medium such as sheepskin \cite{cite11} and/or papyrus.

More effective writing means came about so letters increased and scribes became like human photocopiers. Then the printing press \cite{cite12} meant that more communication could be produced than ever before, and not just facts, figures and important notices, but fiction, stories and drawings could be duplicated faster and more effective than ever before.

We were still limited by distance, so we depended on horse and carriage, boats and trains to get our messages across long distance and over the seas. The telegraph, whether it was flashing light, cable or radio, meant that even the most trivia of messages could now be (mostly) painlessly sent to practically every corner of the globe.

There was little to no abuse of these resources, nothing foolish or time wasting because there was an inherent cost incurred by the sender. You had to pay to send off a letter, you were charged by the telegraph office to send a telegram. It took money, time and manpower to send a message.

Then the internet came along which brought with it easy, fast and extremely efficient communications wherever it could go. This is why email became the killer application of the internet; the one task or operation that made it the biggest thing to happen in the rapid duplication of information in an incredibly short amount of time reaching the highest number of people ever, overtaking the printing press in the list of the most revolutionary milestones mankind has reached in an effort to get information from one point to point in the fastest, cheapest and easiest way.

Unfortunately, the protocol was too open, it was too insecure. It was protected by obscurity; the simple fact that only very few people even knew about it, or could use it and they had to have privileged access to those facilities, which was mostly government and academic related. But that false protection was destroyed by the personal computer with internet access.

Now, messages can be sent by anyone, anywhere, in practically unlimited quantity, to a great number of people, at extremely low cost (in many cases, free), without verification, without consequences from the message's content, at any time. This is the problem that modern networks call "spam". Named for the Monty Python sketch where the dialogue turns from simple and funny dialogue to just the word "spam" repeated over and over, by every member in the cast until there is no meaningful dialogue left.

Spam has become a real problem for computer networks today and many solutions have been sought to address these issues. PGP was taking good steps, but it needed technical proficiency to get it working, which is unsuited for modern users. So XMPP was designed in 1999 to address these problems on a whole new level. It is sender verified, built-in message encryption for secure communications, decentralised and flexible.

However, while XMPP is an ideal protocol for sending messages, it does not work like SMTP does for email. So in order to get people using XMPP like they would POP3 and IMAP, there needs to be a translation layer, or something to aide in the transition between the two messaging protocols. The flexibility of the XMPP might initially be quite alien to a lot of users, so it needs to be presented to them in the most practical and familiar way possible in order to gain a high level of acceptance.

The ideal way would be to have a server with both SMTP and XMPP support, but at the moment this would be a huge and complex task that would need to be designed very specially, considering SMTP is an all-or-nothing protocol and XMPP is much more secure and controlled.

Another way would be have a system that uses the existing email infrastructure to retrieve messages already waiting, but which are prepared in a certain way that could be detected, then it would forward the contents of those marked messages to users on the XMPP network so they would be able to receive messages just like normal, with no disruption to existing services or infrastructure. That is what we look at and how to implement and deploy it.

One day soon, a more integrated solution will be made that assists in the transition even further that takes advantage of an existing SMTP server, but does not use mailboxes at all. This is in the works and will be part of a future paper.

For now, baby steps are needed to raise awareness of this innovative solution to the email security problems and how to get others using it and have it slowly integrated into existing networks so it can be demonstrated and have users reap the benefits of. The problem of binary data i.e:: email attachments is the last major hurdle that needs to be solved, but we must remember that it is solvable and thus a solution could be just around the corner waiting for us.

Communication is a wonderful thing, and with the right protocol that allows near instant messaging, sender and receiver verification, built-in message security, flexibility and a decentralised infrastructure such as XMPP, communication doesn't mean spending hours deleting spam from your inbox and needing to keep complex email filters up to date but instead, enjoying a clean messaging system that does one thing and does it well; get secure messages from one point to another, with the least amount of cost and effort and time.


\bibliographystyle{plain}

\end{document}